\documentclass[aps,twocolumn,groupedaddress,showpacs,showkeys,amsfonts,amssymb,amsmath]{revtex4}
\usepackage{graphics}
\usepackage{ltablex,multirow}
\usepackage[colorlinks=true,linkcolor=blue,citecolor=blue,urlcolor=blue]{hyperref}
\usepackage{soul,color}

\begin{document}

%Title of paper
\title{Phonons in the multiferroic langasite Ba$_3$NbFe$_3$Si$_2$O$_{14}$ : evidences for symmetry breaking}

\author{C. Toulouse}
\affiliation{Laboratoire Mat\'eriaux et Ph\'enom\`enes Quantiques UMR 7162 CNRS, Universit\'e Paris Diderot-Paris 7, 75205 Paris c\'edex 13, France}
\author{M. Cazayous}
\affiliation{Laboratoire Mat\'eriaux et Ph\'enom\`enes Quantiques UMR 7162 CNRS, Universit\'e Paris Diderot-Paris 7, 75205 Paris c\'edex 13, France}
\author{S. de Brion}
\email{sophie.debrion@neel.cnrs.fr}
\affiliation{Universit\'e Grenoble Alpes, Institut N\'eel, F-38000 Grenoble, France}
\affiliation{CNRS, Institut N\'eel, F-38000 Grenoble, France}
\author{F. Levy-Bertrand}
\affiliation{Universit\'e Grenoble Alpes, Institut N\'eel, F-38000 Grenoble, France}
\affiliation{CNRS, Institut N\'eel, F-38000 Grenoble, France}
\author{H.Barkaoui}
\affiliation{Universit\'e Grenoble Alpes, Institut N\'eel, F-38000 Grenoble, France}
\affiliation{CNRS, Institut N\'eel, F-38000 Grenoble, France}
\author{P. Lejay}
\affiliation{Universit\'e Grenoble Alpes, Institut N\'eel, F-38000 Grenoble, France}
\affiliation{CNRS, Institut N\'eel, F-38000 Grenoble, France}
\author{L. Chaix}
\thanks{Present address: Stanford Institute for Materials and Energy Sciences, SLAC National Accelerator Laboratory, Menlo Park, California 94025, USA}
\affiliation{CNRS, Institut N\'eel, F-38000 Grenoble, France}
\affiliation{Universit\'e Grenoble Alpes, Institut N\'eel, 38042 Grenoble, France}
\affiliation{Institut Laue-Langevin, 6 rue Jules Horowitz,38042 Grenoble, France}
\author{M.B. Lepetit}
\affiliation{CNRS, Institut N\'eel, F-38000 Grenoble, France}
\affiliation{Universit\'e Grenoble Alpes, Institut N\'eel, 38042 Grenoble, France}
\affiliation{Institut Laue-Langevin, 6 rue Jules Horowitz,38042 Grenoble, France}
\author{J. B. Brubach}
\affiliation{Synchrotron SOLEIL, L'Orme des Merisiers Saint-Aubin, BP 48, F-91192 Gif-sur-Yvette c\'edex, France}
\author{P. Roy  }
\affiliation{Synchrotron SOLEIL, L'Orme des Merisiers Saint-Aubin, BP 48, F-91192 Gif-sur-Yvette c\'edex, France}

%============================================================================
\begin{abstract}
The chiral langasite Ba$_3$NbFe$_3$Si$_2$O$_{14}$ is a multiferroic compound. While its magnetic order below T$_N$=27~K is now well characterised, its polar order is still controversial. We thus looked at the phonon spectrum and its temperature dependence to unravel possible crystal symmetry breaking. We combined optical measurements (both infrared and Raman spectroscopy) with ab initio calculations and show that signatures of a polar state are clearly present in the phonon spectrum even at room temperature. An additional symmetry lowering occurs below 120~K as seen from emergence of softer phonon modes in the THz range. These results confirm the multiferroic nature of this langasite and open new routes to understand the origin of the polar state.
\end{abstract}

%75.85.+t	Magnetoelectric effects, multiferroics (for multiferroics and magnetoelectric films, see 77.55.Nv)
%78.30.-j	Infrared and Raman spectra
%78.20.Bh	Theory, models, and numerical simulation
%75.25.Dk	Orbital, charge, and other orders, including coupling of these orders
%PACS 71.27.+a  Strongly correlated electron systems


\maketitle
%============================================================================

%\section{Introduction}

%%%%%%%%%%%%%%%%%%%%%%%%%%%%%%%%%%%%%%%%%%%%%%%%%
The Fe langasite Ba$_3$NbFe$_3$Si$_2$O$_{14}$ is a fascinating material owing to its original chiral and magnetic properties~\cite{MAR08,LOI11, SIM12}. Recently, its THz spectrum revealed a new kind of electro magnetic excitation,  exhibiting all the characters of an optical phonon, plus the ability to be excited by the magnetic field of a THz wave~\cite{CHAI13}. A model involving a helical polarisation has been proposed, in which the symmetry of the crystallographic structure  (P321) is reduced. Two transition temperatures are then expected: the first one involving the magnetic order is clearly observed at T$_N$=27~K, the second one involving the establishment of a static polarisation and therefore the loss of crystallographic symmetry remains controversial. While the THz magneto-electric excitation appears below T$_P$=120~K, no structural transition has been reported so far. Moreover, the compound has been shown to sustain a weak static electric polarisation below T$_N$, but studies differ concerning the direction of this polarization~\cite{ZHO09,LEE14}. To unravel possible symmetry breaking in this langasite compound, we  probed potential structural changes by looking at the phonon spectrum using infrared and Raman spectroscopies. We also confronted our experimental results with first principle calculations.

\begin{figure}[h]
\resizebox{7.6cm}{!}{\includegraphics{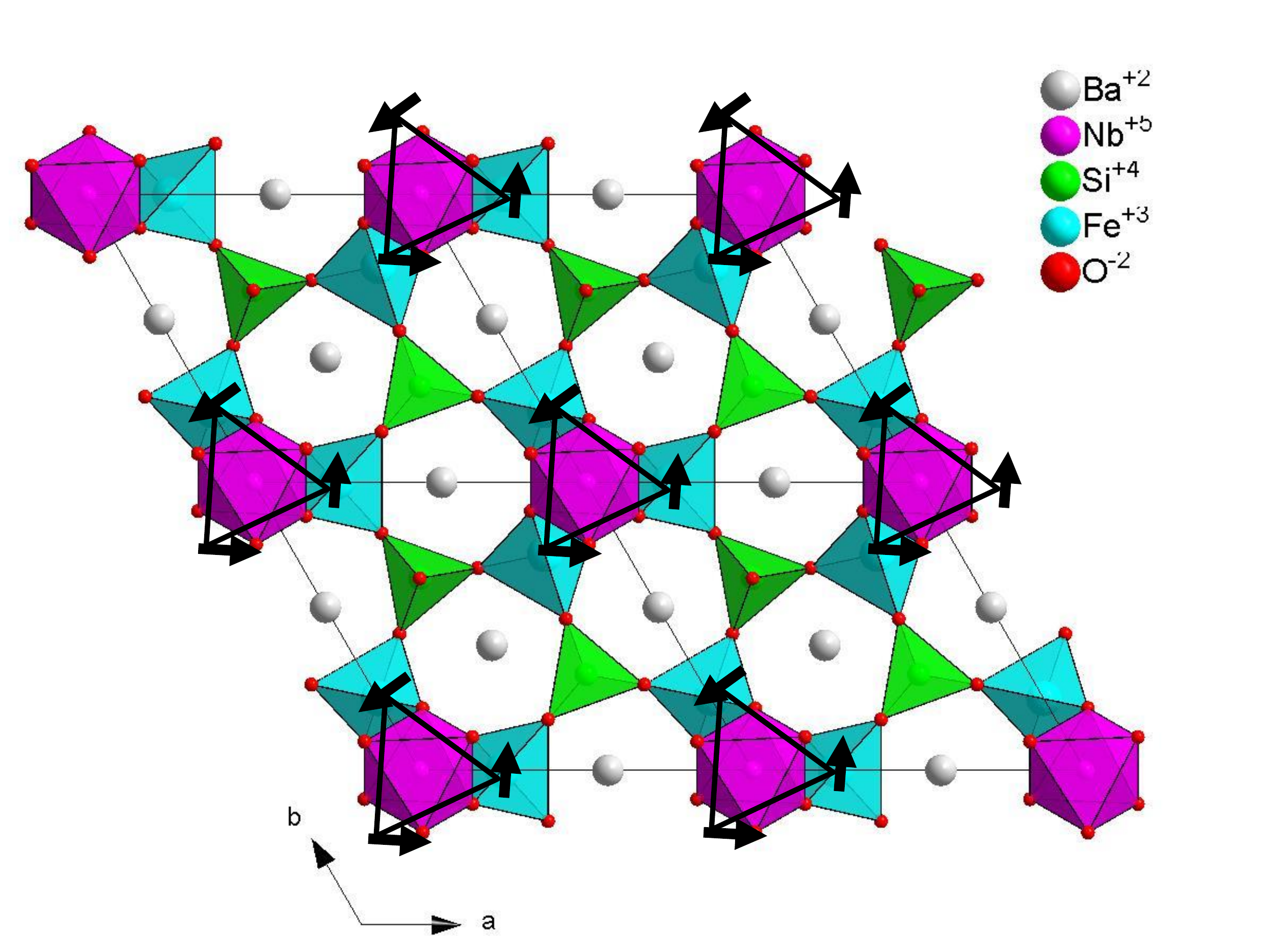}}
\caption{ Ba$_3$NbFe$_3$Si$_2$O$_{14}$  crystallographic structure and magnetic order projected along the $\bf{c}$-axis. $\bf{a,b}$-axes are 2-fold axes while $\bf{c}$-axis is the 3-fold axis.}
\label{fig1}
\end{figure}

Ba$_3$NbFe$_3$Si$_2$O$_{14}$ crystallises in the P321 space group~\cite{MIL98}. The magnetic Fe$^{3+}$ ions form a lattice of triangles arranged in a triangular network stacked along the $\bf{c}$-axis (see Figure~\ref{fig1}) . This space group is non polar and contains one 3-fold axis, the $\bf{c}$-axis, as well as three 2-fold axes, in the $\bf{a,b}$-plane perpendicular to the $\bf{c}$-axis. A static polarisation in the $\bf{a,b}$-plane is present only if a symmetry breaking occurs, most probably  from P321 to C2, with the loss of the three-fold axis. On the other hand, a static polarisation along the $\bf{c}$-axis is not allowed for C2 but possible for P$3$, the other subgroup of P321 where the 3-fold axis is preserved and no 2-fold axis remains. Finally, in the lowest symmetry, P1, a static polarisation in the $\bf{a,b}$-plane as well as along the $\bf{c}$-axis is allowed. Optical measurements should be particularly well suited to probe such a symmetry breaking. For P321 symmetry, the 66 optical modes, expected for this compound with 23 atoms per unit cell, consist in 22 E modes that are doubly degenerated, 10 A$_1$ modes and 12 A$_2$ modes. Infrared (IR) spectroscopy  can probe E and A$_2$ modes while Raman spectroscopy can probe E and A$_1$ modes. In  C2 symmetry as well as in P3 and P1 symmetry, all 66 modes are no longer degenerated and are all IR and Raman active.

\section{Lattice excitations : phonon modes}

\subsection{Infrared measurements}
 Infrared measurements were performed using two kind of experimental  set-up. Temperature resolved data were recorded from 50~cm$^{-1}$  up to 700~cm$^{-1}$  on a powdered sample, using the synchrotron radiation on the AILES beamline at SOLEIL, combined with an IFS 125 spectrometer and a helium cooled bolometer. The sample temperature was scanned from 8 K to 300 K. Spectra were obtained in the transmission configuration with a resolution of 1~cm$^{-1}$. The powdered sample was staked between two polymer films stretched on a copper disk with a 4~mm diameter hole. An identical copper disk with polymer film was measured at room temperature and used as a reference to determine the absolute transmission $T$. We present here absorbance data i.e. $Abs=-LogT$. Single crystal measurements were recorded at room temperature with a vertex 70v spectrometer in a reflection configuration within $10^\circ$ from the normal incidence. The refelctivity was measured in two spectral ranges: in the far infrared range,  50-690~cm$^{-1}$ at 1~cm$^{-1}$ resolution, we used a mercury lamp, a silicon beamsplitter combined with an helium cooled bolometer. In the mid infrared range, 560-6000~cm$^{-1}$ at 2~cm$^{-1}$ resolution, we used a globar lamp, a KBr beamsplitter combined with a nitrogen cooled MCT detector. A gold mirror was used as a reference t odetermine the absolute reflection. Two different plaquettes were measured, with the $\bf{c}$-axis perpendicular to the plaquette plane (Z configuration) or within the plane (Y configuration). Phonon modes with E symmetry are expected in both Z and Y configuration, while A$_2$ symmetry modes should be visible in the Y configuration only.

\begin{figure}
\resizebox{7.6cm}{!}{\includegraphics{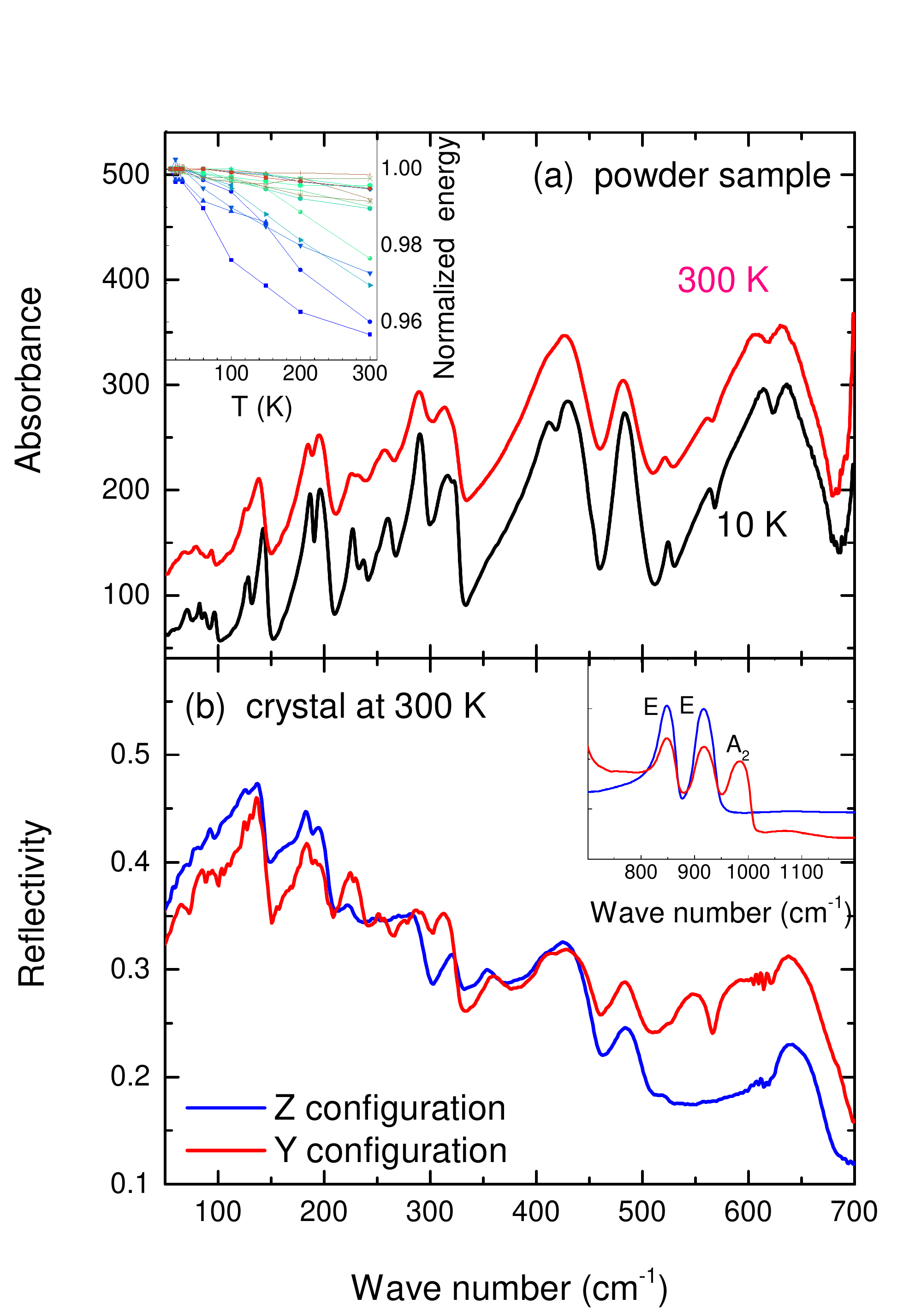}}
\caption{ (a) Infrared Absorbance at 13~K (below T$_N$) and 300~K (above T$_P$) obtained on a powder sample in the energy range 50 to 700~cm$^{-1}$ on the AILES beamline at SOLEIL. Insert: Temperature dependence of phonon energy normalised at 13~K.  (b) Infrared reflectivity at 300~K obtained on two single crystals (Y and Z configurations) in the energy range 50 to 700~cm$^{-1}$ as well as 700 to 1200~cm$^{-1}$ (insert).}
\label{fig2}
\end{figure}

 The infrared spectra from 50~cm$^{-1}$ up to 700~cm$^{-1}$ are presented in Figure~\ref{fig2} for the powder spectra recorded at different temperatures at SOLEIL  as well as for the single crystal spectra at room temperature recorded on the Vertex spectrometer up to 1200~cm$^{-1}$. Many phonon modes are visible extending from 70~cm$^{-1}$ up to 1000~cm$^{-1}$. The room temperature data from both sets of measurements are consistent. The vertex data allow us to identify some A$_2$ and E modes while the synchrotron data give us the temperature dependence of the whole phonon spectra below 700~cm$^{-1}$ with a high sensitivity.  We identified a total of  30 modes among the 34 expected in P321 symmetry. A close inspection on the powder spectra reveals that, the 30 observed total number of modes remains constant in the whole temperature range, which extends from well below the magnetic ordering temperature (27~K) to well above the supposed helical polarisation temperature (120~K). Clearly, no signature of symmetry breaking as a function of temperature is evidenced here.

To further confirm P321 symmetry through the whole temperature range, Raman measurements were performed.

\subsection{Raman measurements}
Those measurements were performed on the single crystal samples used for reflectivity measurements (Y and Z configurations). Spectra were recorded in a backscattering geometry with a triple spectrometer, Jobin Yvon T64000, coupled to a liquid-nitrogen-cooled CCD detector using a Torus 532 solid laser emitting at 532~nm. The high rejection rate of the spectrometer allows to detect low energy excitations down to 5~cm$^{-1}$. Measurements between 7 and 300~K have been performed using an ARS closed-cycle He cryostat.

\begin{figure}
\resizebox{8.6cm}{!}{\includegraphics{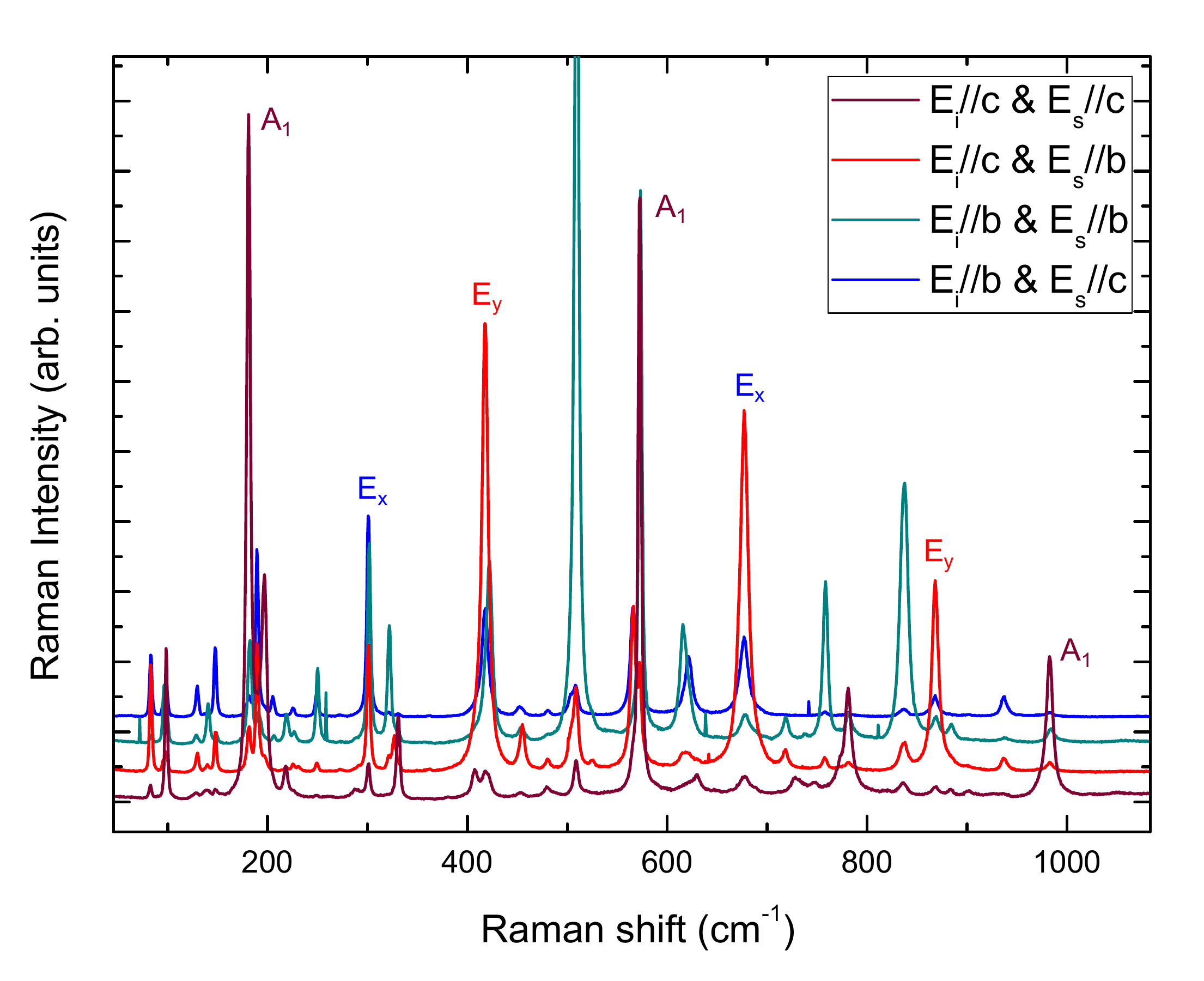}}
\caption{Raman spectra at 15~K obtained from a single crystal (Y configuration) in the energy range 50 to 1100~cm$^{-1}$ with different polarization configurations. $E_{i}$ and $E_{s}$ correspond to the electric field of the incident and scattered light respectively.}
\label{fig3}
\end{figure}

 Measurements under various polarisation for the incident and scattered light allow to distinguish between E and A$_1$ modes (see Figure~\ref{fig3} for the results at 15~K). Namely, we expect 54 Raman active phonon modes~: 44 $E$ (22 $E_{x}$ $\oplus$ 22 $E_{y}$) $\oplus$ 10 $A_{1}$. Calculated selection rules using the Bilbao Crystallographic Server Raman tensors appear in Table~\ref{tab:SR} for the different polarization configuration for the two samples (in the Y and Z configurations). These selection rules allowed to assign the symmetries of the 44 phonon modes observed between 90 and 1000~cm$^{-1}$~: the 10 predicted $A_{1}$ modes and 34 $E$ modes.

 \begin{table}[htbp]
   \centering
   \begin{tabular}{c|c|c}
   \; \; Polarization \; \; &\; \; \; \; Porto \; \; \; \; & \; \; \; \; Selected \; \; \; \; \\
   configuration & Notation & symmetries\\
   \hline
   \hline
  \; & \; & \;\\
   $E_i$//b, $E_s$//b & $X^*(YY)\overline{X}^*$ & A$_1$ + E$_x$ + E$_y$ \\
   $E_i$//b, $E_s$//c & $X^*(YZ)\overline{X}^*$ & E$_x$ \\
   $E_i$//c, $E_s$//b & $X^*(ZY)\overline{X}^*$ & E$_x$ + E$_y$ \\
   $E_i$//c, $E_s$//c & $X^*(ZZ)\overline{X}^*$ & A$_1$ \\

   \end{tabular}
   \caption{P321 Raman selections rules for the Y configuration crystal.}
   \label{tab:SR}
 \end{table}

The assignment of Raman acive phonons are summarised in Table~\ref{table1} together with the infrared modes. Most of the E modes are observed in both sets of measurements as expected, although more modes are evidenced in Raman spectroscopy.

\begin{figure}
\resizebox{8.6cm}{!}{\includegraphics{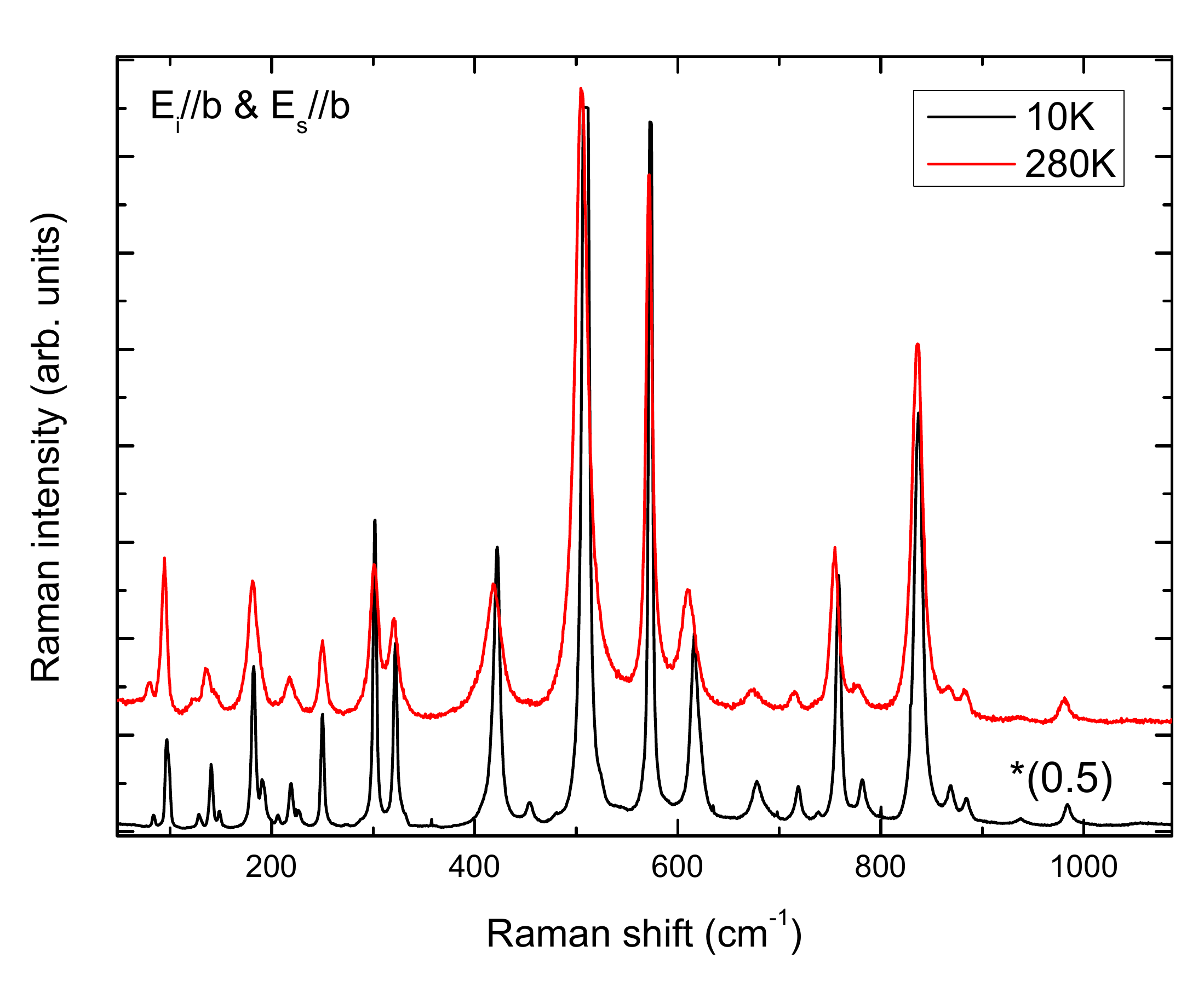}}
\caption{Raman spectra at 10~K (below T$_N$) and 280~K (above T$_P$) obtained on single crystal (Y configuration) in the energy range 50 to 1100~cm$^{-1}$.}
\label{fig4}
\end{figure}

\begin{figure}
\resizebox{8.6cm}{!}{\includegraphics{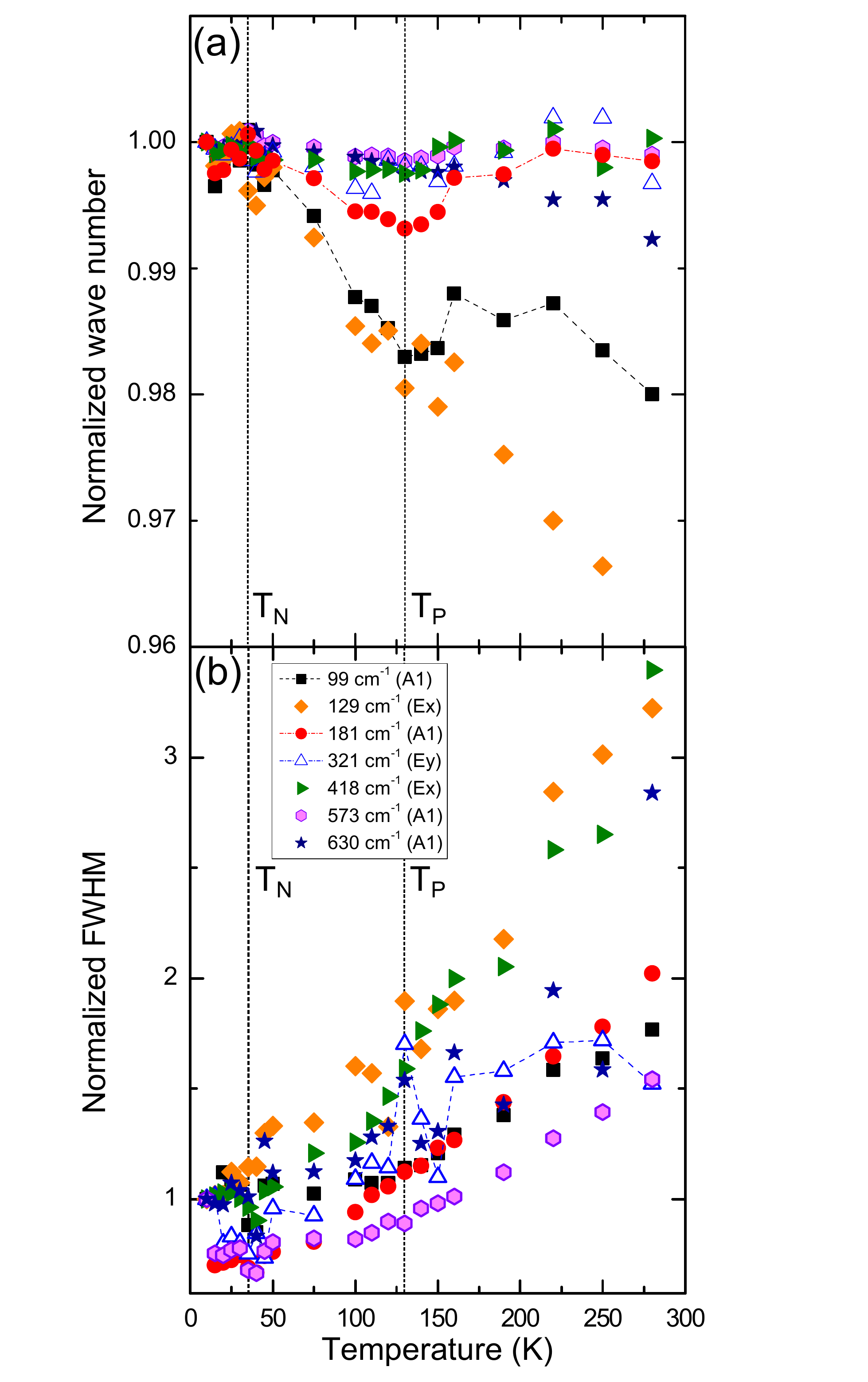}}
\caption{Temperature dependence of the phonon energy (a) and spectral width (b), both normalized by their value at 10~K.}
\label{fig5}
\end{figure}

The temperature dependence of the Raman spectrum has been studied from 10~K up to room temperature. Typical spectra recorded at 10~K (below T$_N$) and 280~K (above T$_P$) are presented in Figure~\ref{fig4}. As observed for the infrared measurements, the number of Raman modes remain unchanged with temperature changes. Figure \ref{fig5}.a shows the energy of the most intense modes normalized by their value at 10~K. Most of them exhibit a hardening around 30~K, indicating a sensitivity to the magnetic transition (T$_N$=~27~K). An additional evolution is clearly visible: several modes present an unusual softening starting at 150~K, followed by a hardening around 120~K. This is particularly evident for modes at 99 and 181~cm$^{-1}$. To complete this observation we have reported on Fig \ref{fig5}.b the change in spectral width of these excitations (using a lorentzian fit) relative to their values at  10~K. As expected, most of the phononic peaks are thinner at low temperature indicating a longer lifetime, except for  the mode at 321 cm$^{-1}$ exhibiting an increase of spectral width around 130~K. The combination of both observations suggests that the compound may undergo a structural transition around 130~K, at T$_P$, when the magneto-electric excitation has been observed.

Let us now compare our experimental data with calculated phonon modes.

\section{First principle calculations}
 We used density functional theory (DFT) with the B3LYP~\cite{B3LYP} hybrid functional to optimize the geometry and compute the phonon spectrum in the P321 space group. Hybrid functionals are known to better treat correlated systems (such as the present case), since the exact exchange part mostly corrects the self interaction problem and the gap underestimation (known to affect LDA and GGA calculations). The calculation was done with the CRYSTAL~\cite{CRYSTAL09} code, using 3$\zeta$+p basis set quality for the Fe, O and Si atoms, small core pseudo-potential and associated basis set for the Nb and Ba ions~\cite{POP}. The optimised geometries for all studied space groups compare in a similar way with the experimental data of reference~\cite{MAR08} with a maximum displacement of 0.03~Å located on the oxygen atoms at position O3. The average displacement is less than 0.005~Å and the degree of lattice distortion is 0.008.

The computed phonons spectrum is reported in Table~\ref{table1} with the experimental assignments in the  P321 space group. Among the 66 calculated modes, only one cannot be assigned (99~cm$^{-1}$). The agreement between optical measurements and calculation seems therefore reasonable  since, except for some of the A2 modes for which the difference between the computed and experimental frequencies are somewhat larger than usually expected for DFT calculations ($\sim$30-40~cm$^{-1}$), all other modes agree within 10~cm$^{-1}$ with experimental ones. There are however additional non assigned Raman modes (see table~\ref{table1}) in addition to the weak modes that may arise from impurities or sample misalignment. Further on, for four observed modes at 273~cm$^{-1}$, 503~cm$^{-1}$, 868~cm$^{-1}$  and 884~cm$^{-1}$ the wrong symmetry is predicted: E modes according to Raman spectroscopy, while in this energy range only A1 or A2 modes are computed: a A2 mode at 280~cm$^{-1}$, a A1 mode at 491~cm$^{-1}$, a A2 mode at 870~cm$^{-1}$  and a A1 mode at 877~cm$^{-1}$.

\begin{table*}[h]
  \begin{minipage}[t]{0.58\linewidth}
    \begin{ruledtabular}
      \begin{tabular}[c]{c@{\hspace*{1ex}}c@{\hspace{5ex}}c@{\hspace*{1ex}}c@{\hspace{5ex}}c@{}c@{\hspace{5ex}}}
        \multicolumn{2}{c}{Raman\hspace*{5ex}} &  \multicolumn{2}{c}{IR\hspace*{5ex}} &
        \multicolumn{2}{l}{Calc. SG: P321\hspace*{5ex}}  \\
        Freq & Irrep &  Freq  & Irrep& Irrep & Freq \\
        \colrule
        . & . & 71 & . & A2 & 71  \\
        83 & E & 82 &   & E  & 81 \\
        . & . & 87 & . & A2 & 84  \\
        96 & E & 96 & . & E  & 92  \\
        99 &A1 & .  & . & A1 & 96   \\
        \multicolumn{4}{c}{\color{blue}\bf No assignation} & {\color{blue}\bf E}  &  {\color{blue}\bf 99} \\      %--
        . & . & 126& . & A2 &125\\
        129& E & 128& . & E  &127 \\
        . & . & 142& .  & A2 &140 \\ %--
        140& E &  . & . & E  &141  \\
        190& E &189 & . & E  &181\\
        181&A1 & .  & . & A1 &183  \\
        197&A1 & .  & . & A1 &193 \\
        206&E  &196 & . & E  &196  \\      %--
        . & . &236 & . & A2 &222 \\
        218&A1 & .  & . & A1 &224     \\
        226& E &227 & . & E  &230 \\
        250& E &250 & . & E  &251  \\
        . & . &260 & . & E  &252  \\
        273&\color{blue}\bf E & .  & . & A2 &280 \\ %--
        weak& . &290 & . & E  &289 \\
        301& E &301 & . & E  &301\\
        321& E &322 & . & E  &322 \\
        331&A1 &    &   & A1 &335  \\
        408& E &weak& . & E  &392    \\
        418& E &411 & . & E  &408   \\
        454& E &weak& E & E  &440  \\
        480& E &484 & E & E  &465\\
        . & . &430 & . & A2 &472\\
        503&\color{blue}\bf E & .  & . & A1 &491  \\ %--
        . & . &546 &A2 & A2 &516  \\
        525& E &524 & E & E  &520  \\
        . & . &598 &A2 & A2 &559 \\
        566& E &565 & . & E  &563\\
        573&A1 & .  & . &A1  &565    \\
        615& E &608 & E & E  &599 \\
        622& E & .  & . & E  &625   \\
        781&A1 & .  & . & A1 &789  \\
        836& E & 847& E & E  &831  \\
        868&\color{blue}\bf  E &    &   &A2  &870    \\ %-
        884&\color{blue}\bf E &    &   &A1  &878     \\ %-
        . & . &917 & E & E  &904\\
        . & . &983 &A2 & A2 &968 \\
        983&A1 & .  & . & A1 &981
      \end{tabular}
    \end{ruledtabular}
  \end{minipage} \hfill
%
% Modes non assignes
%
  \begin{minipage}[t]{0.4\linewidth}
    \begin{ruledtabular}
%      \begin{tabular}[c]{c@{}c@{\hspace{5ex}}c@{}c@{}}
      \begin{tabular}[c]{p{5ex}p{5ex}@{\hspace{8ex}}p{5ex}p{5ex}}
        \multicolumn{4}{c}{Non assigned modes} \\
        \multicolumn{4}{c}{within  P321 calcul.} \\
        \multicolumn{2}{c}{Raman\hspace*{4ex}} &
        \multicolumn{2}{c}{IR\hspace*{0ex}} \\
        Freq & Irrep &  Freq & Irrep \\
        \colrule
%        .  & . &\color{blue}\bf 30 & \color{red}\bf A2  \\
        {\color{blue}\bf 148}& {\color{blue}\bf E} &{\color{blue}\bf weak}& .  \\ %--
        . & . &\color{blue}\bf 316 & .  \\ %--
        \color{blue}\bf 362&\color{blue}\bf E & .  & .  \\ %--
        \color{blue}\bf 422&\color{blue}\bf E & .  & .  \\ %--
        \color{blue}\bf 509&\color{blue}\bf E & .  & .  \\ %--
        \color{blue}\bf 677&\color{blue}\bf E & .  & .  \\ %-
        \color{blue}\bf 630&\color{blue}\bf A1& .  & .  \\ %-
        . & . &\color{blue}\bf 640 &\color{blue}\bf  E  \\ %--
        \color{blue}\bf 677 & \color{blue}\bf E & . & . \\
        \color{blue}\bf 718 & \color{blue}\bf E & . & . \\
%       \color{blue}\bf 737 & . & . & . \\
%       \color{blue}\bf 748 & . & . & . \\
        \color{blue}\bf 758&\color{blue}\bf E & .  & .  \\ %-
        \color{blue}\bf 772 & . & . & . \\
        \color{blue}\bf 937&\color{blue}\bf E & .  & .   %-
      \end{tabular}
    \end{ruledtabular}
  \end{minipage}
  \caption{Measured and calculated phonons modes in the P321 space group. Frequencies are in $\rm cm^{-1}$.}
\label{table1}
\end{table*}

 On this basis,  it seems clear that the P321 space group must be revisited. A close examination at the non assigned experimental modes reveals that most of them are very close to a E assigned one. In fact,in the P321 space group,  $E_{x}$ and $E_{y}$ should be identical and we should only see 10+22=32 different frequencies in Raman scattering and not 44 as observed. Such a splitting of the doubly degenerated E modes prompt us to abandon the 3-fold axis as a  symmetry element. Under this condition the space group should be lowered from P321 to its C2 subgroup. The irreducible representation correspondence between the two groups are as such
  \begin{eqnarray*}
    {\rm P321} && {\rm C2} \\
    A_1 & \rightarrow & A \\
    A_2 & \rightarrow & B \\
    E & \rightarrow & A + B
  \end{eqnarray*}
Table~\ref{tab:phC2} reports the phonon modes computed within the C$_2$ group and their experimental assignment. Note that, within this group, there are two negative phonons frequencies in the B irreducible representation indicating that this group is unstable and the C$_2$ symmetry should also be revisited, at least in the low temperature phase. There are still four computed modes unassigned and a few experimental modes that do not correspond to any calculated frequency (in blue in table~\ref{tab:phC2}). We thus computed again the phonon spectrum within the P$_1$ space group (also reported in table~\ref{tab:phC2}). As expected, all calculated phonon modes are now found stable. In addition all computed modes can easily be assigned to experimental ones with a good accuracy. Indeed the average error between computed and measured frequencies is within 6.1\,cm$^{-1}$. One should however note that the few experimental modes in the 640--770\,cm$^{-1}$ range remain unassigned.

A few conclusions can be drawn at this point. First, the P321 space group usually assumed in the literature for this system is not the correct group for the $\rm  Ba_3NbFe_3Si_2O_{14}$  langasite. Indeed, the $C_3$ rotation around the c-axis is broken, even at room temperature ---~thus allowing a static polarisation in the (a,b) plane to take place~--- but also the in plane $C_2$ rotation axis is expected to be lost, allowing also for a  polarisation along the c direction. Even when lowering the symmetry, the calculations could not account for the phonons in the whole frequency range. Since all the other phonon modes are reproduced with a very good accuracy, it is unlikely for the computational technique to be at fault. One should thus look for other reasons. Crystalline phases impurities may be invoked to explain the extra experimental modes, however some of the excitations observed in Raman spectroscopy within the 640-770~cm$^{-1}$ range are very intense. In addition, the 677 and 772~cm$^{−1}$ modes have previously been measured on a different sample by an other group ~\cite{HUDL}. It is therefore unlikely that impurities could explain the experiment-theory discrepancies. The only remaining possibility is that both the initial point group and the unit cell are incorect.

At this point one should remember that the magnetic unit cell corresponds to a $\rm c'=7c$ super-cell. We therefore computed the phonon modes in a double (2c) and triple (3c) cell along the c-direction within the P321 space group. Unfortunately the size of these calculations does not allow to go up to a 7c unit cell or P1 space group. Nevertheless, these calculations clearly show that even in a double or triple unit cell, we still do not find phonon modes in the 640-770~cm$^{-1}$ range. The spectrum of the triple unit cell however presents the emergence of new phonon modes at very low energy ($< 70~cm^{-1}$).

\begin{table*}[h]
  \begin{minipage}[c]{0.3\textwidth}
%  \begin{ruledtabular}
    \begin{tabularx}{0.3\textwidth}{c@{\hspace*{1ex}}c@{\hspace*{3ex}}
                       p{1ex}@{\hspace*{-2ex}}c@{\hspace{1ex}}c}
\hline \hline \\[-1.8ex]
Raman & IR & \multicolumn{3}{c}{Calc. SG: C2\hspace*{5ex}}  \\
Freq  &Freq&& Irrep & Freq \\
\colrule \\[-1.8ex]
\endhead
\hline \hline
\endfoot
    &     &&  B  & -137  \\
    &     &&  B  &  -17  \\
    &  71 &&  B  &   76  \\
83  &  82 &&  A  &   81  \\
    &  87 &&  A  &   93  \\
99  &     &&  A  &   96  \\
\multirow{2}{*}{96}&\multirow{2}{*}{96} &\multirow{2}{*}{$\left\{\rule{0ex}{2.7ex}{}\right.$}&  A  &   97  \\
    &     &&  B  &  110  \\
    &     &&{\color{blue}\bf B}&{\color{blue}\bf 116}  \\
    & 126 &&  A  &  124  \\
129 & 128 &&  B  &  124  \\
    & 142 &&  B  &  137  \\
140 &     &&  A  &  139  \\
148 & weak&&  B  &  145  \\
    &     &&{\color{blue}\bf B}&{\color{blue}\bf 174}  \\
\multirow{2}{*}{181}&     &\multirow{2}{*}{$\left\{\rule{0ex}{2.7ex}{}\right.$}&  A  &  179  \\
    &     && A  &  180  \\
\multirow{2}{*}{190}&\multirow{2}{*}{189} &\multirow{2}{*}{$\left\{\rule{0ex}{2.7ex}{}\right.$}&  B  &  191  \\
    &     &&  A  &  191  \\
197 & 196 &&  A  &  194  \\
206 &     &&  B  &  201  \\
218/222  &    &&  A  &  223 \\
226 & 227 &&  A  &  228  \\
    & 236 &&  B  &  230  \\
\multirow{3}{*}{250}&\multirow{3}{*}{250}&\multirow{3}{*}{$\left\{\rule{0ex}{4ex}{}\right.$}&  B  &  248  \\
    &     &&  A  &  249  \\
    &     &&  B  &  250  \\
    & 260 &&  B  &  259  \\
273 &     &&  B  &  280  \\
    &\multirow{2}{*}{290}&\multirow{2}{*}{$\left\{\rule{0ex}{2.7ex}{}\right.$}&  A  &  288  \\
    &     &&  B  &  290  \\
301 & 301 &&  A  &  300  \\
    & 316 &&  B  &  316  \\
321 & 322 &&  A  &  320  \\
331 &     &&  A  &  333  \\
362 &     &&  B  &  379  \\
    &     &&{\color{blue}\bf A}&{\color{blue}\bf 391}  \\
408 & 411 &&  B  &  402  \\
418 &     &&  A  &  406  \\
422 &     &&  B  &  419  \\
    & 430 &&  A  &  438  \\
454 &     &&  B  &  450  \\
    &     &&{\color{blue}\bf A}&{\color{blue}\bf 465}  \\
    &     &&{\color{blue}\bf B}&{\color{blue}\bf 469}  \\
480 &     &&  B  &  473  \\
    & 484 &&  A  &  488  \\
503 &     &&  B  &  515  \\
509 &     &&  A  &  517  \\
525 & 524 &&  B  &  519  \\
    & 546 &&  B  &  556  \\
573 &     &&  A  &  561  \\
\multirow{2}{*}{566}&\multirow{2}{*}{565}&\multirow{2}{*}{$\left\{\rule{0ex}{2.7ex}{}\right.$}&  A  &  562  \\
    &     &&  B  &  563  \\
    & 598 &&  B  &  591  \\
615 & 608 &&  A  &  594  \\
622 &     &&  B  &  621  \\
630 &     &&  A  &  623  \\
781 &     &&  A  &  788  \\
836 &     &&  A  &  831  \\
    & 847 &&  B  &  841  \\
868 &     &&  B  &  868  \\
884 &     &&  A  &  876  \\
    &\multirow{2}{*}{917}&\multirow{2}{*}{$\left\{\rule{0ex}{2.7ex}{}\right.$}&  A  &  905  \\
    &     &&  B  &  906  \\
937 &     &&  B  &  965  \\
983 & 983 &&  A  &  978  \\
 \end{tabularx}
%    \end{ruledtabular}
  \end{minipage} \hfill
  \begin{minipage}[c]{0.3\textwidth}
%  \begin{ruledtabular}
    \begin{tabularx}{0.3\textwidth}{c@{\hspace*{1ex}}c@{\hspace*{3ex}}
                       p{1ex}@{\hspace*{-2ex}}c@{\hspace{1ex}}c }
\hline \hline \\[-1.8ex]
Raman & IR & \multicolumn{3}{c}{Calc. SG: P1\hspace*{5ex}}  \\
Freq  &Freq&&Irrep &  Freq   \\
\colrule \\[-1.8ex]
\endhead
\hline \hline
\endfoot
    &  71 && A &  74  \\
\multirow{3}{*}{83} & & \multirow{3}{*}{$\left\{\rule{0ex}{4ex}{}\right.$} & A &  82  \\
    &     &   & A &  83  \\
    &     &   & A &  84  \\
    &\multirow{2}{*}{87}&\multirow{2}{*}{$\left\{\rule{0ex}{2.7ex}{}\right.$}& A & 93  \\
    &     &   & A &  94  \\
\multirow{2}{*}{96}&\multirow{2}{*}{96}&\multirow{2}{*}{$\left\{\rule{0ex}{2.7ex}{}\right.$}&A&  96  \\
    &     && A &  97  \\
 99 &     && A &  98  \\
    & 126 && A & 124  \\
\multirow{2}{*}{129}&\multirow{2}{*}{128}&\multirow{2}{*}{$\left\{\rule{0ex}{2.7ex}{}\right.$}& A &124  \\
    &     && A & 124  \\
140 &     && A & 139  \\
    & 142 && A & 140  \\
148 & weak&& A & 142  \\
\multirow{3}{*}{181}&     &\multirow{3}{*}{$\left\{\rule{0ex}{4ex}{}\right.$} &  A &179  \\
    &     && A & 180  \\
    &     && A & 181  \\
190 & 189 && A & 192  \\
197 & 196 && A & 194  \\
206 &     && A & 196  \\
218 &     && A &  222  \\
222 &     && A & 223  \\
226 & 227 && A & 229  \\
    & 236 && A & 230  \\
\multirow{2}{*}{250}&\multirow{2}{*}{250}&\multirow{2}{*}{$\left\{\rule{0ex}{2.7ex}{}\right.$}& A & 249  \\
    &     &   & A & 250  \\
    & 260 && A & 250  \\
273 &     && A & 279  \\
\multirow{2}{*}{weak}&\multirow{2}{*}{290}&\multirow{2}{*}{$\left\{\rule{0ex}{2.7ex}{}\right.$}& A & 288  \\
    &     &   & A & 288  \\
\multirow{2}{*}{301}&\multirow{2}{*}{301}&\multirow{2}{*}{$\left\{\rule{0ex}{2.7ex}{}\right.$}& A &299  \\
    &     && A & 301  \\
    & 316 && A & 319  \\
321 & 322 && A & 320  \\
331 &     && A & 333  \\
\multirow{2}{*}{362}&     &\multirow{2}{*}{$\left\{\rule{0ex}{2.7ex}{}\right.$}&A & 392  \\
    &     &   & A & 393  \\
408 & 411 && A & 407  \\
418 &     && A & 408  \\
422 &     && A & 439  \\
    & 430 && A & 440  \\
\multirow{2}{*}{454}& & \multirow{2}{*}{$\left\{\rule{0ex}{2.7ex}{}\right.$}& A & 465  \\
    &     && A & 466  \\
480 &     && A & 473  \\
    & 484 && A & 489  \\
503 &     && A & 512  \\
509 &     && A & 517  \\
525 & 524 && A & 517  \\
    & 546 && A & 557  \\
\multirow{2}{*}{566}&\multirow{2}{*}{565}&\multirow{2}{*}{$\left\{\rule{0ex}{2.7ex}{}\right.$}& A &560  \\
    &     && A & 561  \\
573 &     && A & 563  \\
    & 598 && A & 595  \\
615 & 608 && A & 597  \\
622 &     && A & 623  \\
630 &     && A & 625  \\
781 &     && A & 789  \\
836 &     && A & 832  \\
    & 847 && A & 832  \\
868 &     && A & 868  \\
884 &     && A & 875  \\
    &\multirow{2}{*}{917}&\multirow{2}{*}{$\left\{\rule{0ex}{2.7ex}{}\right.$}& A & 905  \\
    &     && A & 905  \\
937 &     && A & 963  \\
983 & 983 && A & 978  \\
 \end{tabularx}
%    \end{ruledtabular}
  \end{minipage} \hfill
 \begin{minipage}[c]{0.3\textwidth}
   \begin{ruledtabular}
     \begin{tabular}[c]{c@{\hspace{1ex}}c}
       \multicolumn{2}{c}{Non assigned modes} \\
        \multicolumn{2}{c}{within  C2 and P1 calcul.} \\
        Raman freq.&
        IR  freq.\\
        \colrule
%        .  & . &\color{blue}\bf 30 & \color{blue}\bf A2  \\
        . & \color{blue}\bf 640   \\ %--
        \color{blue}\bf 677 &  . \\
        \color{blue}\bf 718 &  . \\
%        \color{blue}\bf 737 & . & . & . \\
%        \color{blue}\bf 748 & . & . & . \\
        \color{blue}\bf 758& .  \\ %-
        \color{blue}\bf 772& .  \\
 \end{tabular}
\end{ruledtabular}
 \end{minipage}
    \caption{Calculated (within the C2 and P1 space groups) and measured (IR and Raman) phonons modes.  Frequencies are in $\rm cm^{-1}$.}
    \label{tab:phC2}
  \end{table*}

We will thus now focus on the low energy part of the phonon spectrum, below (70~cm$^{-1}$), the calculated cut off energy for the phonons in a single unit cell.

\section{Low energy excitations}

\subsection{THz measurements}

\begin{figure}
\resizebox{7.6cm}{!}{\includegraphics{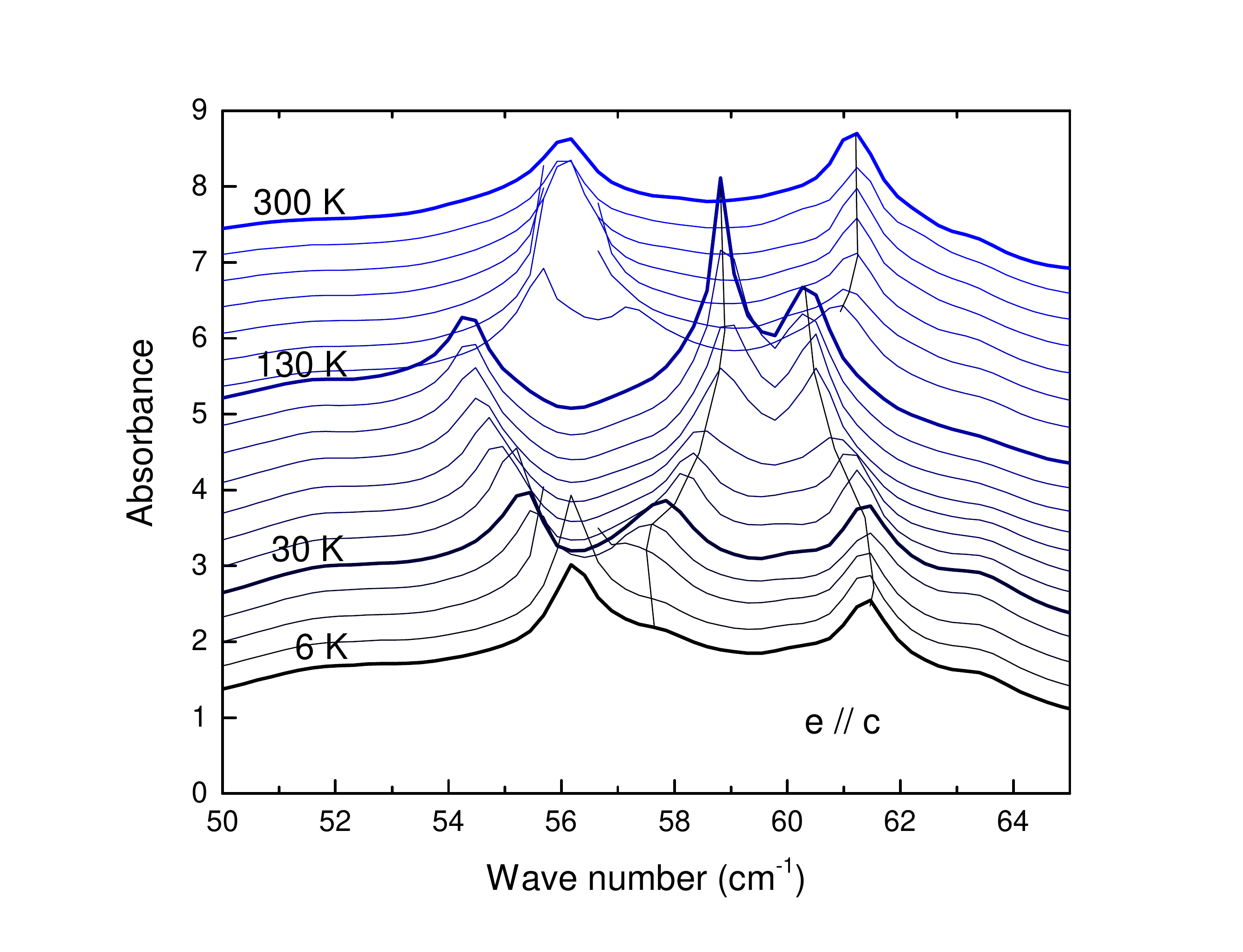}}
\caption{THz spectra for a single crystal in Y configuration in the energy range 50 to 65~cm$^{-1}$ on the AILES beamline at SOLEIL. }
\label{fig6}
\end{figure}

\begin{figure}
\resizebox{7.6cm}{!}{\includegraphics{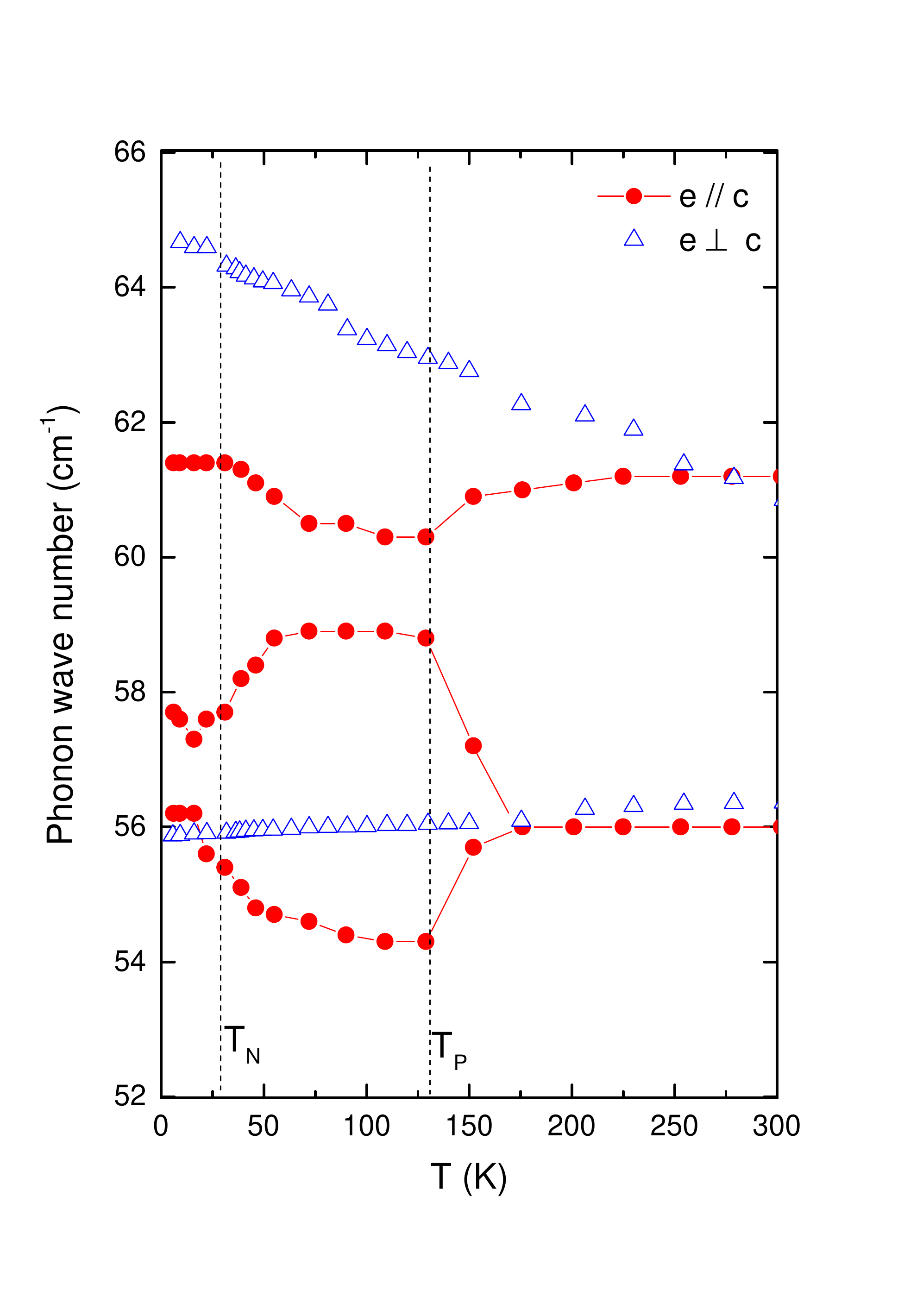}}
\caption{ Phonon energy measured as a function of temperature for two directions of the THz polarisation.}
\label{fig7}
\end{figure}

Extension to the lower energy of the infrared spectra was performed on the AILES beamline at SOLEIL on single crystals using a Helium pumped bolometer with the same experimental conditions as in \cite{CHAI13}. Results are presented in Figure~\ref{fig6}. Sharp modes are observed between 54~cm$^{-1}$ and 62~cm$^{-1}$. Since they are very sharp and their spectral weight is small compared to the phonon modes above 70~cm$^{-1}$, we attribute them to localised defects inside the sample or at its surface. Their temperature dependence is plotted in Figure~\ref{fig7} for two different polarisation of the THz wave. Clearly, the mode at 56~cm$^{-1}$ splits below T$_P$=120~K, an indication that its local environment has lowered its symmetry. Such local defects act  as a probe of crystallographic changes in the bulk crystal that occur below T$_P$.

At even lower energy, as previously published  \cite{CHAI13}, a magneto-electric excitation is observed below T$_P$=120~K. At 16~K for instance, this electro-active excitation occurs at 29~cm$^{-1}$ and the magneto-active one at 23~$cm^{-1}$. Signature of these excitations can also be found in Raman spectroscopy.

\subsection{Raman measurements}

\begin{figure}[h]
\resizebox{8.6cm}{!}{\includegraphics{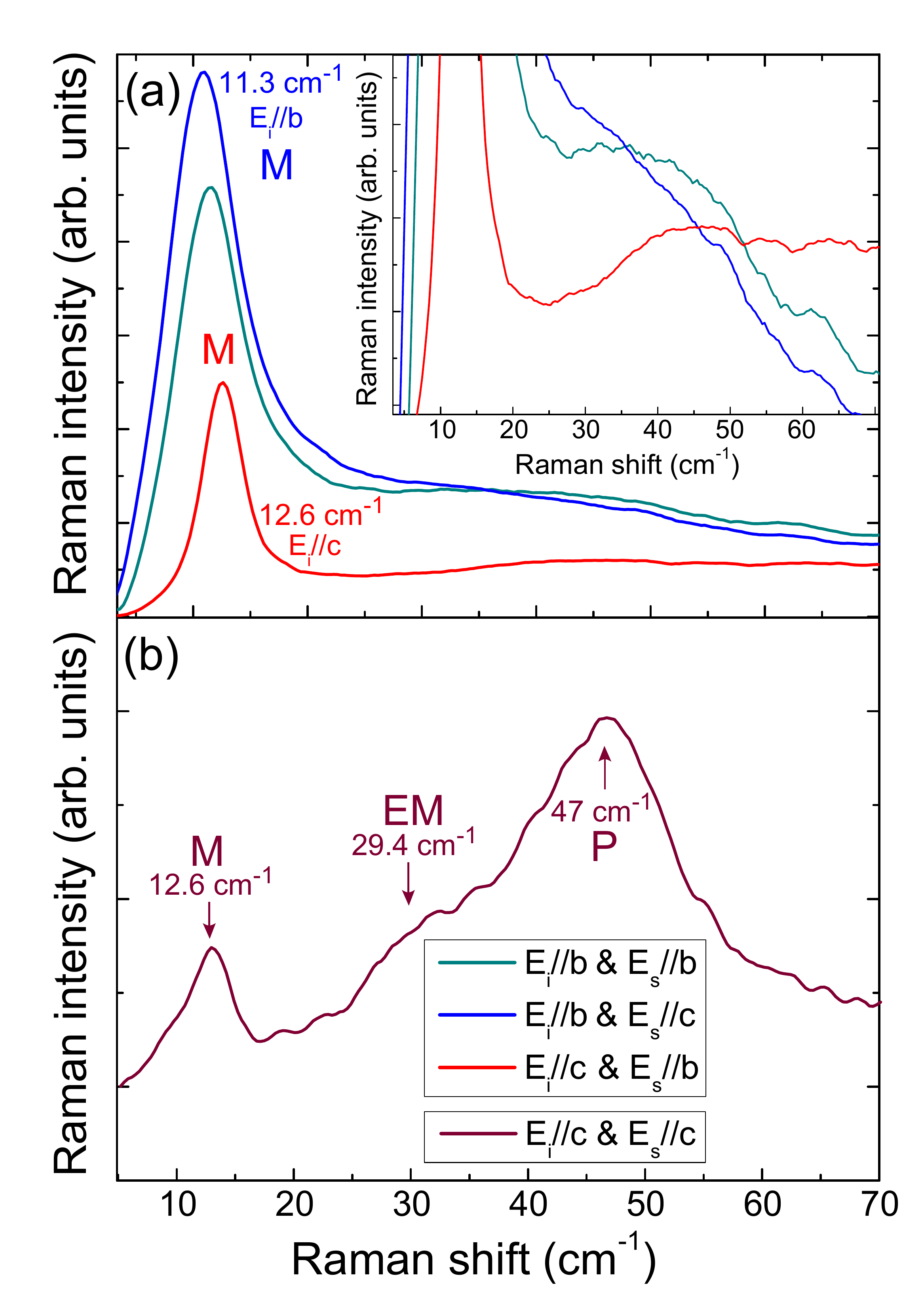}}
\caption{Low energy Raman spectra obtained for different polarization configurations on the Y crystal at 10~K.}
\label{fig8}
\end{figure}

\begin{figure}[h]
\resizebox{8.6cm}{!}{\includegraphics{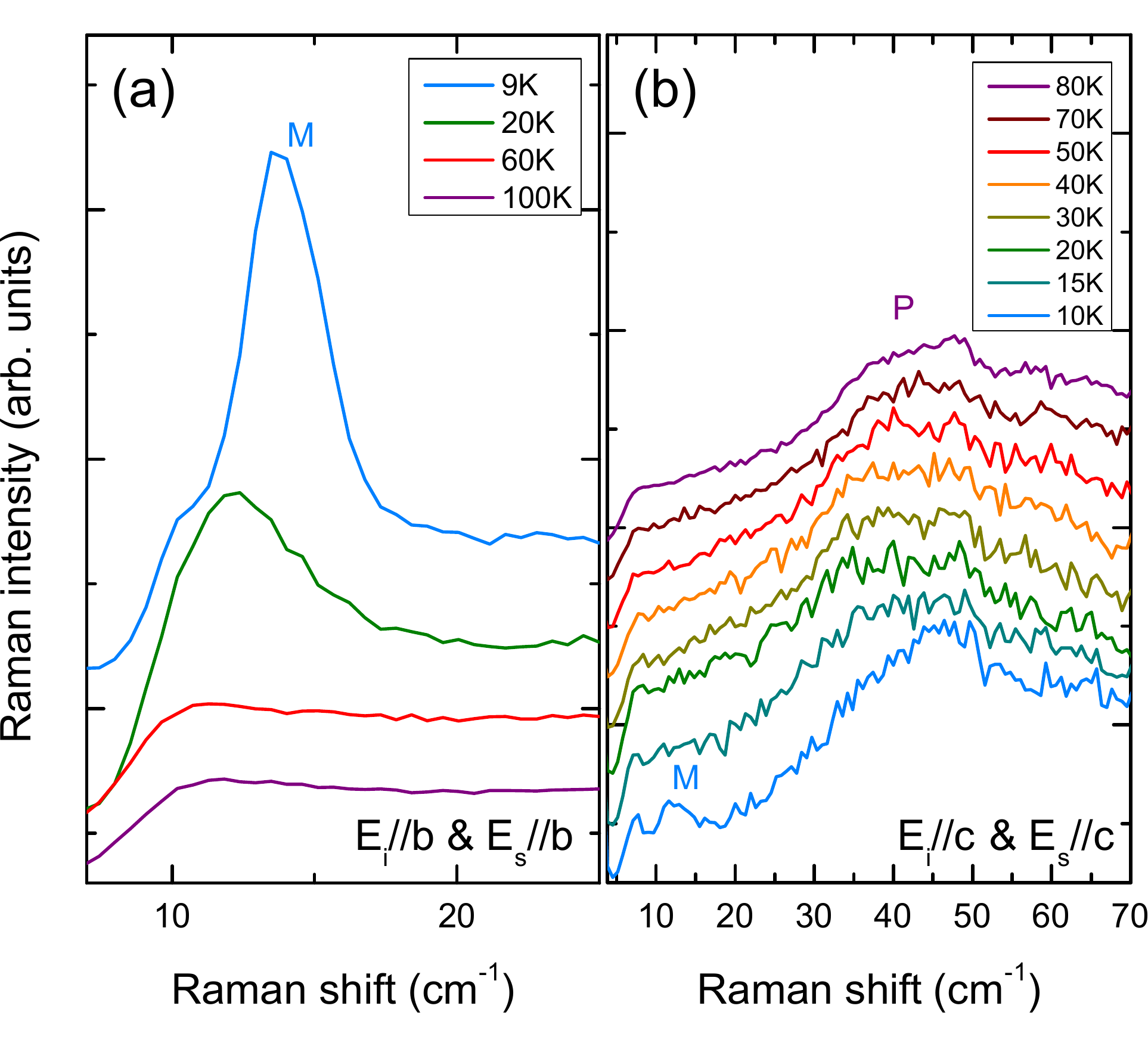}}
\caption{Low energy Raman spectra on the Y sample for several temperatures taken with the electric field of the incident and scattered light respectively along the b-axis (\textbf{a.}) and the c-axis (\textbf{b}).}
\label{fig9}
\end{figure}

Thanks to our Raman optical set-up, we could measure low energy excitations in the 10 to 70~cm$^{-1}$ range. The spectra for all polarization configurations in the Y-crystal (back-scattering along the a-axis) are shown on Fig.\ref{fig8}. To have a better understanding of the physical nature of these low-energy Raman modes, we also varied the temperature between 10 and 100~K, crossing the magnetic transition at T$_N$=~27K. The spectra temperature dependence is reported in Fig.\ref{fig9} for two different polarizations.

At 10~K (Fig.\ref{fig8}), an excitation around 12~cm$^{-1}$, (M), is always present whatever the polarization (with a slight shift), at the same energy as the magnon observed by THz spectroscopy (see Fig.3 in reference \cite{CHAI13}). We therefore assign it to a magnon. This is further confirmed by its temperature dependence (Fig.\ref{fig9})a: as the temperature is increased, it shifts to lower energy and disappears above T$_N$=~27K

Apart from this magnon, other excitations are present: at  29.4~cm$^{-1}$ and 47 cm$^{-1}$ for ($E_i$//c $E_f$//c), and weaker modes are obserevd at around 45~cm$^{-1}$, 50~cm$^{-1}$ and 62~cm$^{-1}$ for the other polarization. The mode at 29.4~cm$^{-1}$, (EM), agrees very well with the electro-magnetic excitation observed in \cite{CHAI13}: at the same temperature, its electro active part is observed at 29~cm$^{-1}$. It has been shown that its position does not change substantially with temperature but it disappears above T$_P$=120~K. %This behavior is confirmed in our Raman measurements: in Fig.\ref{fig9}b, it is clearly observed, with no change of position but a clear decrease in intensity up to 100~K, suggesting a disappearance at higher temperatures.
It is interpreted as a phonon mode associated  to atomic rotations, occuring only below T$_P$=120~K when a symmetry lowering occurs \cite{CHAI13} .

At even higher energy, other excitations are present: at 47~cm~$^{-1}$ for $E_i$//c $E_s$//c, around 40-50cm~$^{-1}$ and 62cm~$^{-1}$ for the other polarizations. Their temperature evolution is very similar to the electro-magnon at 29.4~cm$^{-1}$:   the excitation (P) at 42cm~$^{-1}$ for instance (see Fig.\ref{fig9}b) remains at the same position up to 100~K while its intensity decreases substantially.
From all these low energy measurements, we confirm that, below T$_P$=120~K, a symmetry breaking occurs, with several soft phonon modes emerging in the 25- 65~cm$^{-1}$ energy range. This symmetry breaking is further evidenced through one narrow phonon mode, attributed to an impurity, responding to the local symmetry breaking and splits below T$_P$.

\section{Discussion and conclusion}

From the comparison between IR, Raman and theoretical phonons spectra, its is clear  that the P321 space group assumed from X-ray scattering experiments~\cite{MAR08,ZHO09} should be questioned. Our work shows that the most probable space group is P1, even if the symmetry breaking is expected to be weak. Indeed, in the P321 space group, not only many observed modes (13) cannot be assigned to the computed ones, but four modes (at 273, 503, 868 and 884~cm$^{-1}$) detected in Raman scattering as belonging to the the $E$ irreducible representation can only be assigned to $A_1$ or $A_2$ computed modes. These discrepancies lead us to abandon the 3-fold axis,and use rather C2 or P1 symmetries. The appearance of computed negative frequencies in the C2 space group suggests that the C2 symmetry is unstable and further symmetry breaking should take place, towards P1 symmetry. Nevertheless, for a large number of modes, the P321 selection rules are nearly respected, showing that the symmetry breaking remains weak. Indeed, using the Bilbao Crystallographic server one finds  the  maximum distance between the P321 and P1 groups to be only 0.0030\AA{} and the computed structural lattice distortion estimated at 0.0001.

 Nevertheless, this point group symmetry lowering is clearly not sufficient to account for all the experimental observations: five modes in the 640-770~cm$^{-1}$ range remain   unaccounted for within a simple unit cell. Since we excluded an impurity phase as the origin of these modes, one should thus break the translational symmetry and enlarge the unit cell. Calculated phonons spectra in super-cells along the c direction do not however exhibit any sign of new modes in the desired energy range. The only remaining possibility is thus that these modes originate from an enlargement of the unit cell in the (a,b) plane, an enlargement  present even at room temperature.

At temperatures lower than T$_P=120$~K, we see additional low energy ($< 70 \rm cm^{-1}$) features in the phonons spectra suggesting a  symmetry lowering. First principle calculations in triple-cells along the c axis exhibit new modes in this energy range. %(around 60~cm$^-1}$).
This result put into perspective with the fact that the magnetic unit under T$_N$ is found to be a septuple unit cell along the c axis allow us to propose the following scenario: (i) at room temperature the system crystallises in the P1 group with an enlarged unit cell in the (a,b) plane. (ii) At T$_P$ it undergoes a phase transition further increasing the unit cell along the c axis, most probably with an incommensurate vector. Then (iii) at the magnetic transition the latter locks to c'=7c.

In this scenario, the compound can sustain a polarisation, even at room temperature, both in the (a,b) plane and along the c direction. Clearly this polarisation is expected to be faint and difficult to measure,  as outlined by the weakness of the predicted symmetry breaking. Since electric domains are likely present, the observed polarisation may vary from one sample to the other. At lower temperature, when the magnetic order sets in, a coupling mechanism may enhanced the static polarisation that is further enhanced by a static magnetic field~\cite{CHAI15}.

Finally one should point out that the loss of symmetry at room temperature implies that the polar state is not induced by the set in of the magnetic order. The consequence is that  such crystals can no more be considered as magnetically induced multiferroics and thus the origin of their magneto-electric coupling should be revisited.

\begin{acknowledgments}
We acknowledge J. Debray, J. Balay and A. Hadj-Azzem for the powder and crystals preparation. S. de B. and L. C. aknowledge fruitful discussions with R .Ballou and V. Simonet. This work was supported by the French National Research Agency through projects ANR- DYMAGE, ANR-SUBRISSYME, and ANR-DYMMOS as well as the French General Directorate for Armament (DGA). First principle calculations were done at the IDRIS and CRIHAN computed centers under projects number 91842 and 2007013.
\end{acknowledgments}

%%%%%%% References

\end{document}